# A genuine maximally seven-qubit entangled state

Xin-Wei Zha, Hai-Yang Song, Jian-Xia Qi, Dong Wang, Qian Lan

School of Science, Xi'an University of Posts and Telecommunications, Xi'an, 710121, P R China

**C**ontrary to A.Borras et al.'s [1] conjecture, a genuine maximally seven-qubit entangled state is presented. We find a seven-qubit state whose marginal density matrices for subsystems of 1,2-qubits are all completely mixed and for subsystems of 3-qubits is almost completely mixed.

**1. Introduction**

Quantum entanglement is a key element for applications of quantum communications and quantum information. In particular, the entanglement can be used as a quantum resource to carry out a number of computational and information processing tasks. Such tasks include teleportation of an unknown quantum state . Analysis of such quantum phenomena may provide us a better understanding of the structure of the quantum mechanics framework.

Due to its great relevance, from both the theoretical and the practical lpoints of view, it is imperative to explore and characterize all aspects of the quantum entanglement of multipartite quantum systems. A considerable amount of research has recently been devoted to the study of multi-qubit entanglement measures defined as the sum of bipartite entanglement measures over all the possible bi-partitions of the full system[1-22].

The characterization of multipartite entanglement is no simple matter. For four-qubit systems, some maximally entangled state were given by Ref. [2,3]. Recently, we give the criterion for four-qubit state and present some new maximally entangled four-qubit state[23] .

The maximally entangled five-qubit state is disovered by D. K. Brown [4]. Using the numerical search procedure, Borras et al [1] found a kind of six-qubit maximally entangled state. However, for seven-qubit state systems, there are not maximally entangled states which have been reported.  A.Borras et al [1] conjectures that there is no pure state of seven qubits whose marginal density matrices for subsystems of 1,2,or 3 qubits are all completely mixed.

In order to search of maximally seven-qubit entangled state, let us recall the notion of maximally multipartite entangled states (MMES) of n qubits. Paolo Facchi et al [5] define a

MMES as a minimizer of what they shall call the potential of multipartite entanglement,

$$\pi_{ME} = \binom{n}{n_A}^{-1} \sum \pi_A \qquad (1)$$

Where $\pi_A = Tr_A \rho_A^2$, $\rho_A = Tr_{\bar{A}} |\psi\rangle\langle\psi|$, $\pi_A = \pi_{\bar{A}}$. (2)

$Tr_X$ denotes the partial trace over subsystem X.

The quantity $\pi_{ME}$ measure the average bipartite entanglement over all possible balanced bipartition, i.e.

$$\frac{1}{N_A} \leq \pi_{ME} \leq 1 \qquad (3)$$

Where $N_A = 2^{n_A}$.

Without any loss of generality, we first consider a general seven-qubit state

$$\begin{aligned}|\psi\rangle_{1234567} = (&a_0|0000000\rangle + a_1|0000001\rangle + a_2|0000010\rangle + a_3|0000011\rangle \\&+ a_4|0000100\rangle + a_5|0000101\rangle + a_6|0000110\rangle + a_7|0000111\rangle \\&+ a_8|0001000\rangle + a_9|0001001\rangle + a_{10}|0001010\rangle + a_{11}|0001011\rangle \\&+ a_{12}|0001100\rangle + a_{13}|0001101\rangle + a_{14}|0001110\rangle + a_{15}|0001111\rangle \\&+ a_{16}|0010000\rangle + a_{17}|0010001\rangle + a_{18}|0010010\rangle + a_{19}|0010011\rangle \\&+ a_{20}|0010100\rangle + a_{21}|0010101\rangle + a_{22}|0010110\rangle + a_{23}|0010111\rangle \\&+ a_{24}|0011000\rangle + \cdots + \\&+ a_{120}|1111000\rangle + a_{121}|1111001\rangle + a_{122}|1111010\rangle + a_{123}|1111011\rangle \\&+ a_{124}|1111100\rangle + a_{125}|1111101\rangle + a_{126}|1111110\rangle + a_{127}|1111111\rangle)_{1234567}.\end{aligned}$$

(4)

and it is assumed that the wave function satisfies the normalization condition $\sum_{i=0}^{127} |a_i|^2 = 1$.

For the seven-qubit state, $\pi_{123} = Tr_{123}\rho_{123}^2$, $\rho_{123} = Tr_{4567} |\varphi\rangle_{1234567} {}_{1234567}\langle\varphi|$,

$$\pi_{124} = Tr_{124}\rho_{124}^2, \quad \rho_{124} = Tr_{3567} |\varphi\rangle_{1234567} {}_{1234567}\langle\varphi|,$$

$$\cdots, \qquad (5)$$

$$\pi_{567} = Tr_{567}\rho_{567}^2, \quad \rho_{567} = Tr_{1234} |\varphi\rangle_{1234567} {}_{1234567}\langle\varphi|,$$

Then we have

$$\pi_{ME} = \frac{1}{35}(\pi_{123} + \pi_{124} + \pi_{125} + \pi_{126} + \pi_{127}$$
$$+ \pi_{134} + \pi_{135} + \pi_{136} + \pi_{137}$$
$$+ \cdots +$$
$$+ \pi_{567}) \qquad (6)$$

From Eqs.(4) and (5), the expression of $\pi_{13}, \pi_{13} \cdots, \pi_{67}, \pi_{123}, \pi_{124} \cdots, \pi_{567}$, $\pi_1, \pi_2 \cdots, \pi_7$ can be obtained, (see Appendix A, B, C).

In order to search of maximally seven-qubit entangled state, let us first consider the following state

$$\begin{aligned}|\psi\rangle_{1234567} = (&a_0|0000000\rangle + a_3|0000011\rangle + a_{13}|0001101\rangle + a_{14}|0001110\rangle \\ &+ a_{17}|0010001\rangle + a_{18}|0010010\rangle + a_{28}|0011100\rangle + a_{31}|0011111\rangle \\ &+ a_{37}|0100101\rangle + a_{38}|0001001\rangle + a_{40}|0101000\rangle + a_{43}|0101011\rangle \\ &+ a_{52}|0110100\rangle + a_{55}|0110111\rangle + a_{57}|0111001\rangle + a_{58}|0111010\rangle \\ &+ a_{68}|1000100\rangle + a_{71}|1000111\rangle + a_{73}|1001001\rangle + a_{74}|1001010\rangle \\ &+ a_{85}|1010101\rangle + a_{86}|1010110\rangle + a_{88}|1011000\rangle + a_{91}|1011011\rangle \\ &+ a_{97}|1100001\rangle + a_{98}|1100010\rangle + a_{108}|1101100\rangle + a_{111}|1101111\rangle \\ &+ a_{112}|1110000\rangle + a_{115}|1110011\rangle + a_{125}|1111101\rangle + a_{126}|1111110\rangle)_{1234567}.\end{aligned} \qquad (7)$$

According to Appendix A, we have

$$\begin{aligned}\pi_{123} &= Tr_{123}\rho_{123}^2 \\ &= \left(|a_0|^2 + |a_3|^2 + |a_{13}|^2 + |a_{14}|^2\right)^2 + \left(|a_{17}|^2 + |a_{18}|^2 + |a_{28}|^2 + |a_{31}|^2\right)^2 \\ &+ \left(|a_{37}|^2 + |a_{38}|^2 + |a_{40}|^2 + |a_{43}|^2\right)^2 + \left(|a_{52}|^2 + |a_{55}|^2 + |a_{57}|^2 + |a_{58}|^2\right)^2 \\ &+ \left(|a_{68}|^2 + |a_{71}|^2 + |a_{73}|^2 + |a_{74}|^2\right)^2 + \left(|a_{85}|^2 + |a_{86}|^2 + |a_{88}|^2 + |a_{91}|^2\right)^2 \\ &+ \left(|a_{97}|^2 + |a_{98}|^2 + |a_{108}|^2 + |a_{111}|^2\right)^2 + \left(|a_{112}|^2 + |a_{115}|^2 + |a_{125}|^2 + |a_{126}|^2\right)^2 \\ &+ 2\left|a_0 a_{112}^* + a_3 a_{115}^* + a_{13} a_{125}^* + a_{14} a_{126}^*\right|^2 + 2\left|a_{17} a_{97}^* + a_{18} a_{98}^* + a_{28} a_{108}^* + a_{31} a_{111}^*\right|^2 \\ &+ 2\left|a_{37} a_{85}^* + a_{38} a_{86}^* + a_{40} a_{88}^* + a_{43} a_{91}^*\right|^2 + 2\left|a_{52} a_{68}^* + a_{55} a_{71}^* + a_{57} a_{73}^* + a_{58} a_{74}^*\right|^2 \end{aligned} \qquad (8)$$

$$\pi_{127} = Tr_{127}\rho_{127}^2$$

$$= \left(|a_0|^2 + |a_{18}|^2 + |a_{28}|^2 + |a_{14}|^2\right)^2 + \left(|a_{17}|^2 + |a_3|^2 + |a_{13}|^2 + |a_{31}|^2\right)^2$$

$$+ \left(|a_{52}|^2 + |a_{38}|^2 + |a_{40}|^2 + |a_{58}|^2\right)^2 + \left(|a_{37}|^2 + |a_{55}|^2 + |a_{57}|^2 + |a_{43}|^2\right)^2$$

$$+ \left(|a_{68}|^2 + |a_{86}|^2 + |a_{88}|^2 + |a_{74}|^2\right)^2 + \left(|a_{85}|^2 + |a_{71}|^2 + |a_{73}|^2 + |a_{91}|^2\right)^2$$

$$+ \left(|a_{112}|^2 + |a_{98}|^2 + |a_{108}|^2 + |a_{126}|^2\right)^2 + \left(|a_{97}|^2 + |a_{115}|^2 + |a_{125}|^2 + |a_{111}|^2\right)^2$$

$$+ 2\left|a_0 a_{97}^* + a_{18} a_{115}^* + a_{28} a_{125}^* + a_{14} a_{111}^*\right|^2 + 2\left|a_{17} a_{112}^* + a_3 a_{98}^* + a_{13} a_{108}^* + a_{31} a_{126}^*\right|^2$$

$$+ 2\left|a_{52} a_{85}^* + a_{38} a_{71}^* + a_{40} a_{73}^* + a_{58} a_{91}^*\right|^2 + 2\left|a_{37} a_{68}^* + a_{55} a_{86}^* + a_{57} a_{88}^* + a_{43} a_{74}^*\right|^2 \quad (9)$$

From Eqs.(1-3) and Appendix , through a great deal of mathematics and theoretical calculations, we can find, if

$$a_0 = a_3 = a_{13} = a_{14} = a_{17} = a_{28} = a_{40} = a_{43} = a_{52} = a_{58} = a_{73}$$
$$= a_{74} = a_{85} = a_{91} = a_{97} = a_{98} = a_{108} = a_{111} = a_{112} = a_{125} = \frac{1}{4\sqrt{2}},$$

$$a_{18} = a_{31} = a_{37} = a_{38} = a_{55} = a_{57} = a_{68}$$
$$= a_{71} = a_{86} = a_{88} = a_{115} = a_{126} = -\frac{1}{4\sqrt{2}} \quad (10)$$

we have

$$|\psi_M\rangle_{1234567} = \frac{1}{4\sqrt{2}}[(|0000000\rangle + |0000011\rangle + |0001101\rangle + |0001110\rangle)$$
$$+ (|0010001\rangle - |0010010\rangle + |0011100\rangle - |0011111\rangle)$$
$$+ (-|0100101\rangle - |0100110\rangle + |0101000\rangle + |0101011\rangle)$$
$$+ (|0110100\rangle - |0110111\rangle - |0111001\rangle + |0111010\rangle) \quad (11)$$
$$+ (-|1000100\rangle - |1000111\rangle + |1001001\rangle + |1001010\rangle)$$
$$+ (|1010101\rangle - |1010110\rangle - |1011000\rangle + |1011011\rangle)$$
$$+ (|1100001\rangle + |1100010\rangle + |1101100\rangle + |1101111\rangle)$$
$$+ (|1110000\rangle - |1110011\rangle + |1111101\rangle - |1111110\rangle)$$

and

$$\pi_{ijk} = Tr_{ijk}\rho_{ijk}^2 = \frac{1}{8}, ijk = 123, 124, \cdots, 567; ijk \neq 127, 457, 367;$$

$$\pi_{127} = Tr_{127}\rho_{127}^2 = \frac{1}{4}, \pi_{457} = Tr_{457}\rho_{457}^2 = \frac{1}{4}, \pi_{367} = Tr_{367}\rho_{367}^2 = \frac{1}{4}, \quad (12)$$

Then, we have

$$\pi_{ME} = \frac{1}{35}(\pi_{123}+\pi_{124}+\pi_{125}+\pi_{126}+\pi_{127}+\pi_{134}+\pi_{135}+\pi_{136}+\pi_{137}$$
$$+\cdots+\pi_{567})=\frac{19}{140} \tag{13}$$

According to Appendix B, it is easy to obtain

$$\pi_{ij} = Tr_{ij}\rho_{ij}^2 = \frac{1}{4}, ij=12,13,\cdots,57,67. \tag{14}$$

Similarly, from Appendix C, we have

$$\pi_i = Tr_i\rho_i^2 = \frac{1}{2}, i=1,2,3,4,5,6,7 \tag{15}$$

Therefore, for the seven-qubit state of Eq.(15), the marginal density matrices for subsystems of 1,2, are all completely mixed.

In summary, according to the definition of Paolo Facchi's, we have presented a maximally entangled seven-qubit state. Borras et al.[1] conjectured that there is no pure state of seven qubits whose marginal density matrices for subsystems of 1,2,or 3 qubits are all completely mixed. But here the seven qubits whose marginal density matrices for subsystems of 1,2, are all completely mixed and for subsystems 3 qubits is almost completely mixed. In this paper, we just started to explore the study of maximally seven-qubit entangled states, we hope this can help deeper comprehension of the complex structure of quantum correlations arising in multi- -qubit entangled states.

**Acknowledgements**

This work is supported bythe National Natural Science Foundation of China (Grant No. 10902083) and Shaanxi Natural Science Foundation under Contract No. 2009JM1007.

References

**References**

[1] A.Borras；A R Plastino；J. Batle；C. Zander；M Casas；A Plastino, J. Phys. A: Math. Gen. 40 (2007) 13407.

[2] A. Higuchi and A. Sudbery, Phys. Lett. A 273, (2000) 213

[3]　Y. Yeo and W. K. Chua, Phys. Rev. Lett. 96, (2006) 060502

[4]I. D. K. Brown, S. Stepney, A. Sudbery and S. L. Braunstein,　J. Phys. A: Math. Gen. 38 (2005) 1119

[5] P.Facchi, G.Florio, G. Parisi and S.Pascazio,Phys.Rev.A 77, (2008) 060304(R)

[6] Lin Chen and Masahito Hayashi，Phys. Rev. A **83**, 022331 (2011)

[7] B. Kraus, Phys. Rev. A **82**, 032121 (2010)

[8] H. T. Cui, Di Yuan, and J. L. Tian, Phys. Rev. A **82**, 062116 (2010)

[9] P. Facchi, G. Florio, S. Pascazio, and F. Pepe, Phys. Rev. A **82**, 042313 (2010)

[10] P. Facchi, G. Florio, C. Lupo, S. Mancini, and S. Pascazio, Phys. Rev. A **80**, 062311 (2009)

[11] C. Kruszynska and B. Kraus, Phys. Rev. A **79**, 052304 (2009)

[12] P. Krammer, H.Kampermann, D. Bruß, R. A. Bertlmann, L. C. Kwek, and C. Macchiavello, Phys. Rev. Lett. **103**, 100502 (2009)

[13] A. S. M. Hassan and P. S. Joag, Phys. Rev. A **80**, 042302 (2009)

[14] S. Tamaryan, T. C. Wei and D. K. Park, Phys. Rev. A. 80, 052315 (2009).

[15] M. Aolita, F. Mintert, Phys. Rev. Lett. 97, 050501 (2006).

[16] Y.S. Weinstein and C.S.Hellberg, Phys.Rev.Lett. 95 (2005)030501.

[17] A.R.R.Carvalho, F.Mintert and A.Buchleitner, Phys.Rev.Lett. 93 (2004)230501.

[18] M.Aolita and F.Mintert, Phys.Rev.Lett. 97 (2006)050501.

[19] J.Casalmiglia etal., Phys.Rev.Lett. 95 (2005)180502.

[20] Scott A J 2004 Phys.Rev.A 69 052330

[21] A. R. R. Carvalho, F. Mintert, and A. Buchleitner,. Phys. Rev. Lett. 93, 230501 (2004).

[22] A.Monras, G.Adesso, S.M.Giampaolo, G.Gualdi, G.B.Davies, and F.Illuminati,. Phys.Rev.A 84 012301 (2011).

[23] Xin-wei Zha, Hai-yang Song, Feng Feng. arXiv: quant-ph/ 1012.0371

Appendix A

From Eq.(4), we can obtain

$$\pi_{123} = Tr_{123}\rho_{123}^2$$
$$= \left(|a_0|^2 + |a_1|^2 + \cdots + |a_{14}|^2 + |a_{15}|^2\right)^2 + \left(|a_{16}|^2 + |a_{17}|^2 + \cdots + |a_{30}|^2 + |a_{31}|^2\right)^2$$
$$+ \left(|a_{32}|^2 + |a_{33}|^2 + \cdots + |a_{46}|^2 + |a_{47}|^2\right)^2 + \left(|a_{48}|^2 + |a_{49}|^2 + \cdots + |a_{62}|^2 + |a_{63}|^2\right)^2$$
$$+ \left(|a_{64}|^2 + |a_{65}|^2 + \cdots + |a_{78}|^2 + |a_{79}|^2\right)^2 + \left(|a_{80}|^2 + |a_{81}|^2 + \cdots + |a_{94}|^2 + |a_{95}|^2\right)^2$$
$$+ \left(|a_{96}|^2 + |a_{97}|^2 + \cdots + |a_{110}|^2 + |a_{111}|^2\right)^2 + \left(|a_{112}|^2 + |a_{113}|^2 + \cdots + |a_{126}|^2 + |a_{127}|^2\right)^2$$
$$+ 2\left|a_0 a_{16}^* + a_1 a_{17}^* + \cdots + a_{15} a_{31}^*\right|^2 + 2\left|a_0 a_{32}^* + a_1 a_{33}^* + \cdots + a_{15} a_{47}^*\right|^2$$
$$+ 2\left|a_0 a_{48}^* + a_1 a_{49}^* + \cdots + a_{15} a_{63}^*\right|^2 + 2\left|a_0 a_{64}^* + a_1 a_{65}^* + \cdots + a_{15} a_{79}^*\right|^2$$
$$+ 2\left|a_0 a_{80}^* + a_1 a_{81}^* + \cdots + a_{15} a_{95}^*\right|^2 + 2\left|a_0 a_{96}^* + a_1 a_{97}^* + \cdots + a_{15} a_{111}^*\right|^2$$
$$+ 2\left|a_0 a_{112}^* + a_1 a_{113}^* + \cdots + a_{15} a_{127}^*\right|^2 + 2\left|a_{16} a_{32}^* + a_{17} a_{33}^* + \cdots + a_{31} a_{47}^*\right|^2$$
$$+ 2\left|a_{16} a_{48}^* + a_{17} a_{49}^* + \cdots + a_{31} a_{63}^*\right|^2 + 2\left|a_{16} a_{64}^* + a_{17} a_{65}^* + \cdots + a_{31} a_{79}^*\right|^2$$
$$+ 2\left|a_{16} a_{80}^* + a_{17} a_{81}^* + \cdots + a_{31} a_{95}^*\right|^2 + 2\left|a_{16} a_{96}^* + a_{17} a_{97}^* + \cdots + a_{31} a_{111}^*\right|^2$$
$$+ 2\left|a_{16} a_{112}^* + a_{17} a_{113}^* + \cdots + a_{31} a_{127}^*\right|^2 + 2\left|a_{32} a_{48}^* + a_{33} a_{49}^* + \cdots + a_{47} a_{63}^*\right|^2$$
$$+ 2\left|a_{32} a_{64}^* + a_{33} a_{65}^* + \cdots + a_{47} a_{79}^*\right|^2 + 2\left|a_{32} a_{80}^* + a_{33} a_{81}^* + \cdots + a_{47} a_{95}^*\right|^2$$
$$+ 2\left|a_{32} a_{96}^* + a_{33} a_{97}^* + \cdots + a_{47} a_{111}^*\right|^2 + 2\left|a_{32} a_{112}^* + a_{33} a_{113}^* + \cdots + a_{47} a_{127}^*\right|^2$$
$$+ 2\left|a_{48} a_{64}^* + a_{49} a_{65}^* + \cdots + a_{63} a_{79}^*\right|^2 + 2\left|a_{48} a_{80}^* + a_{49} a_{81}^* + \cdots + a_{63} a_{95}^*\right|^2$$
$$+ 2\left|a_{48} a_{96}^* + a_{49} a_{97}^* + \cdots + a_{63} a_{111}^*\right|^2 + 2\left|a_{48} a_{112}^* + a_{49} a_{113}^* + \cdots + a_{63} a_{127}^*\right|^2$$
$$+ 2\left|a_{64} a_{80}^* + a_{65} a_{81}^* + \cdots + a_{79} a_{95}^*\right|^2 + 2\left|a_{64} a_{96}^* + a_{65} a_{97}^* + \cdots + a_{79} a_{111}^*\right|^2$$
$$+ 2\left|a_{64} a_{112}^* + a_{65} a_{113}^* + \cdots + a_{79} a_{127}^*\right|^2 + 2\left|a_{80} a_{96}^* + a_{81} a_{97}^* + \cdots + a_{95} a_{111}^*\right|^2$$
$$+ 2\left|a_{80} a_{112}^* + a_{81} a_{113}^* + \cdots + a_{95} a_{127}^*\right|^2 + 2\left|a_{96} a_{112}^* + a_{97} a_{113}^* + \cdots + a_{111} a_{127}^*\right|^2$$

Similarly, we can obtain $\pi_{124}, \pi_{125}, \cdots, \pi_{567}$

Appendix B

From Eq.(4), we can also obtain

$$\pi_{12} = Tr_{12}\rho_{12}^2$$
$$= \left(|a_0|^2 + |a_1|^2 + \cdots + |a_{31}|^2\right)^2 + \left(|a_{32}|^2 + |a_{33}|^2 + \cdots + |a_{63}|^2\right)^2$$
$$+ \left(|a_{64}|^2 + |a_{65}|^2 + \cdots + |a_{95}|^2\right)^2 + \left(|a_{96}|^2 + |a_{97}|^2 + \cdots + |a_{127}|^2\right)^2$$
$$+ 2\left|a_0 a_{32}^* + a_1 a_{33}^* + \cdots + a_{31} a_{63}^*\right|^2 + 2\left|a_0 a_{64}^* + a_1 a_{65}^* + \cdots + a_{31} a_{95}^*\right|^2$$
$$+ 2\left|a_0 a_{96}^* + a_1 a_{97}^* + \cdots + a_{31} a_{127}^*\right|^2 + 2\left|a_{32} a_{64}^* + a_{33} a_{65}^* + \cdots + a_{63} a_{95}^*\right|^2$$
$$+ 2\left|a_{32} a_{96}^* + a_{33} a_{97}^* + \cdots + a_{63} a_{127}^*\right|^2 + 2\left|a_{64} a_{96}^* + a_{65} a_{97}^* + \cdots + a_{95} a_{127}^*\right|^2,$$

Similarly, we can obtain $\pi_{13}, \pi_{14}, \cdots, \pi_{67}$

Appendix C

Furthr, from Eq.(4), it is easy to obtain

$$\pi_1 = Tr_1\rho_1^2$$
$$= \left(|a_0|^2 + |a_1|^2 + |a_3|^2 + |a_4|^2 \cdots + |a_{62}|^2 + |a_{63}|^2\right)^2$$
$$+ \left(|a_{64}|^2 + |a_{65}|^2 + |a_{65}|^2 + |a_{66}|^2 \cdots |a_{126}|^2 + |a_{127}|^2\right)^2$$
$$+ 2\left|a_0 a_{64}^* + a_1 a_{65}^* + \cdots + a_{62} a_{126}^* + a_{63} a_{127}^*\right|^2$$

Certainly, it is easy to obtain $\pi_2, \pi_3, \cdots, \pi_7$